\documentclass[11pt]{article}
\usepackage{amssymb}

\usepackage{moriond,epsfig}

\def\be{\begin{equation}}
\def\ee{\end{equation}}
\def\bea{\begin{eqnarray}}
\def\eea{\end{eqnarray}}

\input{tcilatex}

\begin{document}

\title{INFLUENCE OF MAGNETIC\ FIELD ON\ EFFECTIVE\ ELECTRON-ELECTRON
INTERACTIONS\ IN\ A\ COPPER\ WIRE}
\author{\underline{A. Anthore}, F. Pierre*, H. Pothier, D. Esteve, and M.
H. Devoret}
\address{Service de Physique de l'Etat Condens\'e, Commissariat \`a l'Energie Atomique,
Saclay, \\F-91191 Gif-sur-Yvette, France\\
(* Present address: Michigan State University, East Lansing, MI 48824, USA)}

\maketitle

\abstracts{
We have measured in a copper wire the energy exchange rate between
quasiparticles as a function of the applied magnetic field. We find that
the effective electron-electron interaction is strongly modified by the magnetic field,
 suggesting that magnetic impurities play a role on the interaction processes.}

\vspace*{25mm}

In metallic wires, electron-electron interactions lead to quasi-elastic
scattering, which modifies the electron phase-coherence time, and to inelastic
scattering, which redistributes energy between electrons. Several recent
experiments have shown that these properties can be sample-dependent~\cite%
{Echternach,WLGoteborg}. We have found furthermore that the low-temperature
saturation of the phase-coherence time $\tau _{\varphi }$ of quasiparticles,
observed in several experiments, is correlated with an anomalous dependence
on the exchanged energy $\varepsilon $ of $K(\varepsilon )$, a function proportional to 
the averaged squared interaction
between two quasiparticles. The theoretical predictions,
for pure Coulomb interactions~\cite{AA}, $\tau _{\varphi }\propto T^{-2/3}$
and $K(\varepsilon )\propto \varepsilon ^{-3/2}$, were observed in silver
wires~\cite{relaxAg}, whereas a ``saturating'' $\tau _{\varphi }$ and $%
K(\varepsilon )\propto \varepsilon ^{-2}$ were systematically found in
copper wires~\cite{relax,Fred}. In gold wires in which the presence of a
large concentration of magnetic impurities ($\sim 50$~\textrm{ppm} of iron)
could be assessed, $K(\varepsilon )$ was also found to behave as $%
\varepsilon ^{-2}$, with a particulary large prefactor~\cite{Pesc}. On the
theoretical side, it was recently understood that scattering by magnetic
impurities, which limits the phase coherence time on a large range of
temperature, can also provide another channel for energy exchange between
quasiparticles~\cite{Glazman,georg,Kroha}. In order to probe if magnetic
impurities indeed play a role in copper wires, we have measured the magnetic
field dependence of the energy exchange rate.

\section{Measurement set-up}

\subsection{Principle of the experiment}

In order to measure the energy exchange rates between quasiparticles, we
have determined the local electron energy distribution function $f(x,E)$ in
a stationary out-of-equilibrium situation~\cite{relax}, as described in
Fig.~1. A mesoscopic metallic wire is placed between two large electrodes
implementing Landauer reservoirs. In the absence of interactions, the
population of quasiparticles at a given energy interpolates linearly between
the distribution functions in the contacts, leading, if $k_{B}T\ll eU,$ to a
double-step-shaped distribution function. In the opposite ``hot electron''
regime, \textit{i.e.} when the typical interaction time is much shorter than
the diffusion time $\tau _{D}=L^{2}/D,$ equilibrium is reached locally at
each position along the wire: the energy distribution function $f(x,E)$ is a
Fermi function, with a temperature dependent on the position along the wire~%
\cite{Steinbach,Kozub}. Our experiments focus on the intermediate regime, in
which interactions lead to a significant redistribution of the energy
between quasiparticles, but not to a complete thermalization.

\begin{figure}[t]
\epsfysize=53mm \centerline{\epsfbox{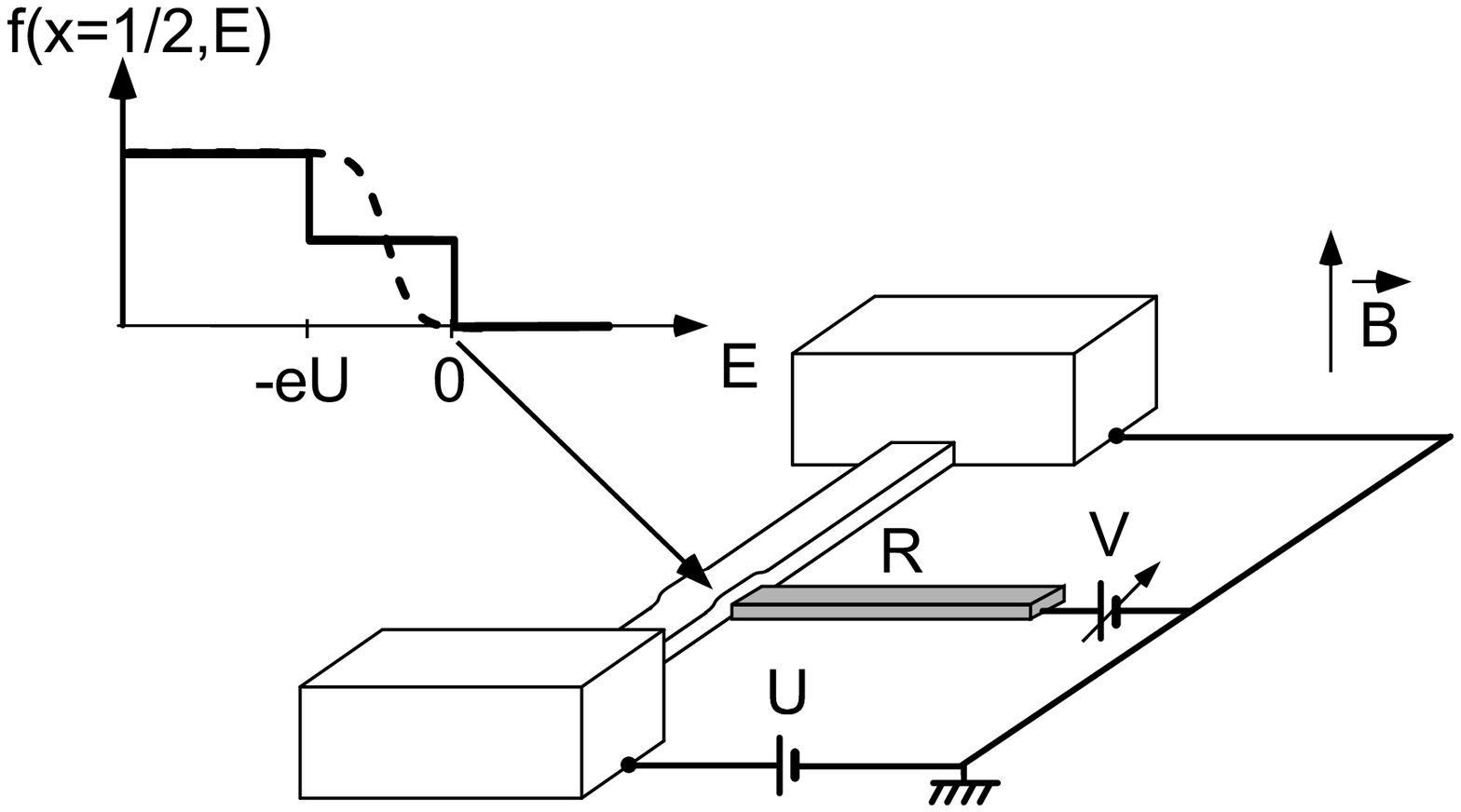}}
\caption{Experimental layout: A metallic wire of length $L$ is connected to
large reservoir electrodes, biased at potentials $0$ and $U$. In the absence
of interactions, the distribution function in the middle of the wire
displays a step with height $1/2$ for energies between $-eU$ and $0$ (solid
curve). When interactions are strong enough to thermalize electrons, the
distribution function is a Fermi function (dashed curve). In the experiment,
the distribution function is obtained from the differential conductance $%
\frac{dI}{dV}(V)$ of the tunnel junction formed by the wire and a reference
resistive electrode placed underneath.}
\end{figure}
In previous experiments~\cite{relaxAg,relax}, the distribution function was
infered from the differential conductance $\frac{dI}{dV}\left( V\right) $ of
a tunnel junction between the wire and a superconducting probe electrode.
Here the probe electrode, also fabricated with a superconducting metal
(Aluminum), was designed as a long ($25~\mathrm{%
%TCIMACRO{\U{b5}}%
%BeginExpansion
{\mu}%
%EndExpansion
m}$), narrow ($150~\mathrm{nm}$) and thin ($12~\mathrm{nm}$) wire in order to present an appreciable
resistance ($\sim 1~\mathrm{k\Omega }$) in the normal state. In
zero-magnetic field, the probe electrode was superconducting and the
distribution function could be obtained as in the earlier experiments. In a
magnetic field larger than $0.15~\mathrm{T}$, the probe electrode was normal
and resistive. In this regime, dynamical Coulomb blockade of tunneling
through the junction resulting from this ``environmental'' resistance~\cite{Devoret} can be
exploited as a quasiparticle energy probe. The relationship between the
differential conductance of the junction $\frac{dI}{dV}\left( V\right) $ and
the distribution function in the wire $f(E)$ is then:%
\begin{equation}
\frac{dI}{dV}(V)=\frac{1}{R_{\mathrm{T}}}\int dE~f(E)~\int d\varepsilon
~P(\varepsilon )~\frac{\partial }{\partial E}((f_{\mathrm{ref}%
}(E+eV+\varepsilon )-f_{\mathrm{ref}}(E+eV-\varepsilon ))  \tag{1}
\label{convolution1}
\end{equation}%
where $R_{\mathrm{T}}$ is the tunnel resistance of the junction, $f_{\mathrm{%
ref}}(E)$ the distribution function in the resistive electrode, assumed to
be a Fermi function at temperature $T_\mathrm{ref}$, and $P(\varepsilon )$ the
probability for an electron to tunnel through the barrier while giving the
environnement the energy $\varepsilon $:%
\[
P(\varepsilon )=\int dt~\exp [i\varepsilon t/\hbar ]~\exp J(t) 
\]%
\newline
with

\[
J(t)=2\int_{0}^{\infty }\frac{d\omega }{\omega }~\frac{\func{Re}[Z_{\mathrm{%
env}}(\omega )]}{R_{K}~}~\frac{\exp [-i\omega t]-1}{1-\exp [-\hbar \omega
/k_{B}T]} 
\]
where $R_{K}=\frac{h}{e^{2}}\approx 25.8~\mathrm{k\Omega }$, $Z_{\mathrm{%
env}}$ and $T$ are the environmental impedance and temperature. 
In our set-up, the environment
consists of the parallel combination of the capacitance $C$ of the junction
and the resistance $R$ of the probe electrode.

In order to evaluate the sensitivity of this novel technique, we show in
Fig. 2 the calculated differential conductance for characteristics of the
environment close to the experimental ones and for two distribution
functions in the wire: a Fermi function, as expected at $U=0$,~and a typical
double-step function for $U=0.3~\mathrm{mV}$. 
The zero-bias conductance dip splits into a double-dip for a double-step
distribution function. In the experiment, we invert the transformation shown
in Fig. 2 and determine the unknown $f(E)$ from the measured $\frac{dI}{dV}%
\left( V\right) $.
\begin{figure}[h]
\epsfxsize=135mm \centerline{\epsfbox{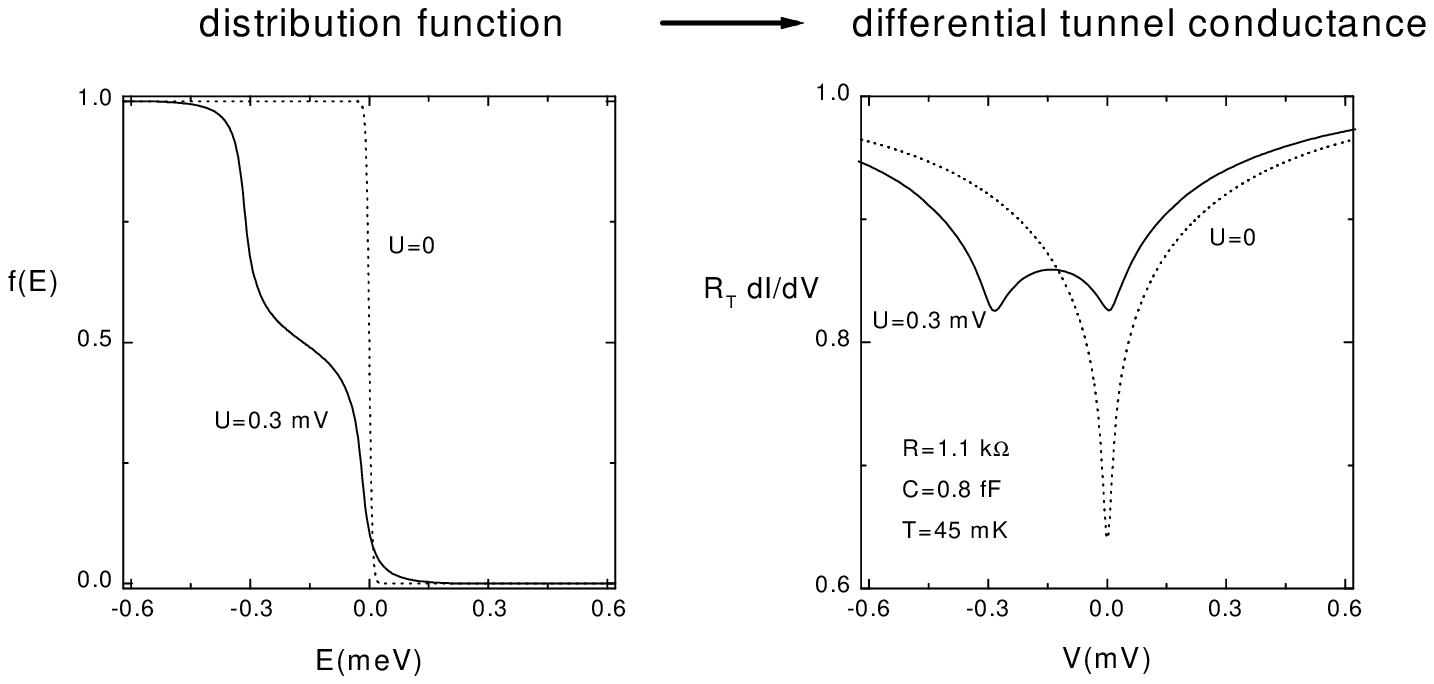}}
\caption{Computer-generated distribution functions at $U=0~$and at $U=0.3~%
\mathrm{mV,}$ with $kT=0.004~\mathrm{meV}$ (left panel), and corresponding
calculated differential conductance of the probe junction (right panel) for
characteristics of the environment close to the experimental ones. }
\end{figure}

\subsection{\protect\bigskip Experimental techniques}

The sample was fabricated in a single pump-down, using three-angle shadow-mask technique through a PMMA suspended mask patterned using e-beam lithography. 
The substrate was thermally oxidized silicon. 
We first deposited a $12~\mathrm{nm}$-thick aluminum film defining the probe
finger. It was then oxidized, in order to obtain the tunnel barrier. The
wire and the reservoirs were obtained by the subsequent evaporation from a $%
99.999~\mathrm{\%}$ purity copper target, at a pressure of $10^{-6}~\mathrm{%
mb}$, at a deposition rate of $1~\mathrm{nm/s}$. The thickness and width of
the wire are $45~\mathrm{nm}$ and $105$~$\mathrm{nm}$. The reservoirs are $%
400~\mathrm{nm}$-thick, with an area of about $1~\mathrm{mm}^{2}$. From the
low temperature wire resistance $R=29.5~\Omega ,$ we deduced from Einstein's
relation, assuming a rectangular cross-section, the diffusion constant $D=90~%
\mathrm{cm}^{2}/\mathrm{s}$ and the diffusion time $\tau _{D}=2.8~\mathrm{ns.%
}$ The sample was mounted in a copper box thermally anchored to the mixing
chamber of a dilution refrigerator. Measurements were performed at $25%
\mathrm{~mK}$. Electrical connections were made through filtered coaxial
lines~\cite{filtres}.

Measurements proceed as follows. In a first step, the voltage $U$ is set
to zero. The measured differential
conductance of the tunnel junction $\frac{dI}{dV}\left( V\right) $, shown in Fig. 3, was found to be independent 
of the magnetic field at $B>0.15~\mathrm{T}$, a value at which all traces of superconductivity in the probe electrode are
washed out.  From the fit of $\frac{dI}{dV}\left( V\right) $ with Eq.~\ref{convolution1}, 
assuming $f(E)=f_\mathrm{ref}(E)$ and $T_\mathrm{ref}=T$, we deduce the tunnel junction conductance 
$R_{\mathrm{T}}^{-1}=23~$\textrm{%
%TCIMACRO{\U{b5}}%
%BeginExpansion
$\mu$%
%EndExpansion
S}, the resistance of the probe electrode 
$R=1.1~$\textrm{k}$\Omega $, the tunnel junction capacitance $C=0.8~\mathrm{fF}$, 
and the temperature $T=68~$\textrm{mK.} This excess temperature compared to the temperature of the mixing chamber 
of the dilution refrigerator (25~mK), 
might be due to an oversimplification of the environmental
impedance, or to external noise. This question still has to be settled. 

\begin{figure}[h]
\epsfysize=58mm \centerline{\epsfbox{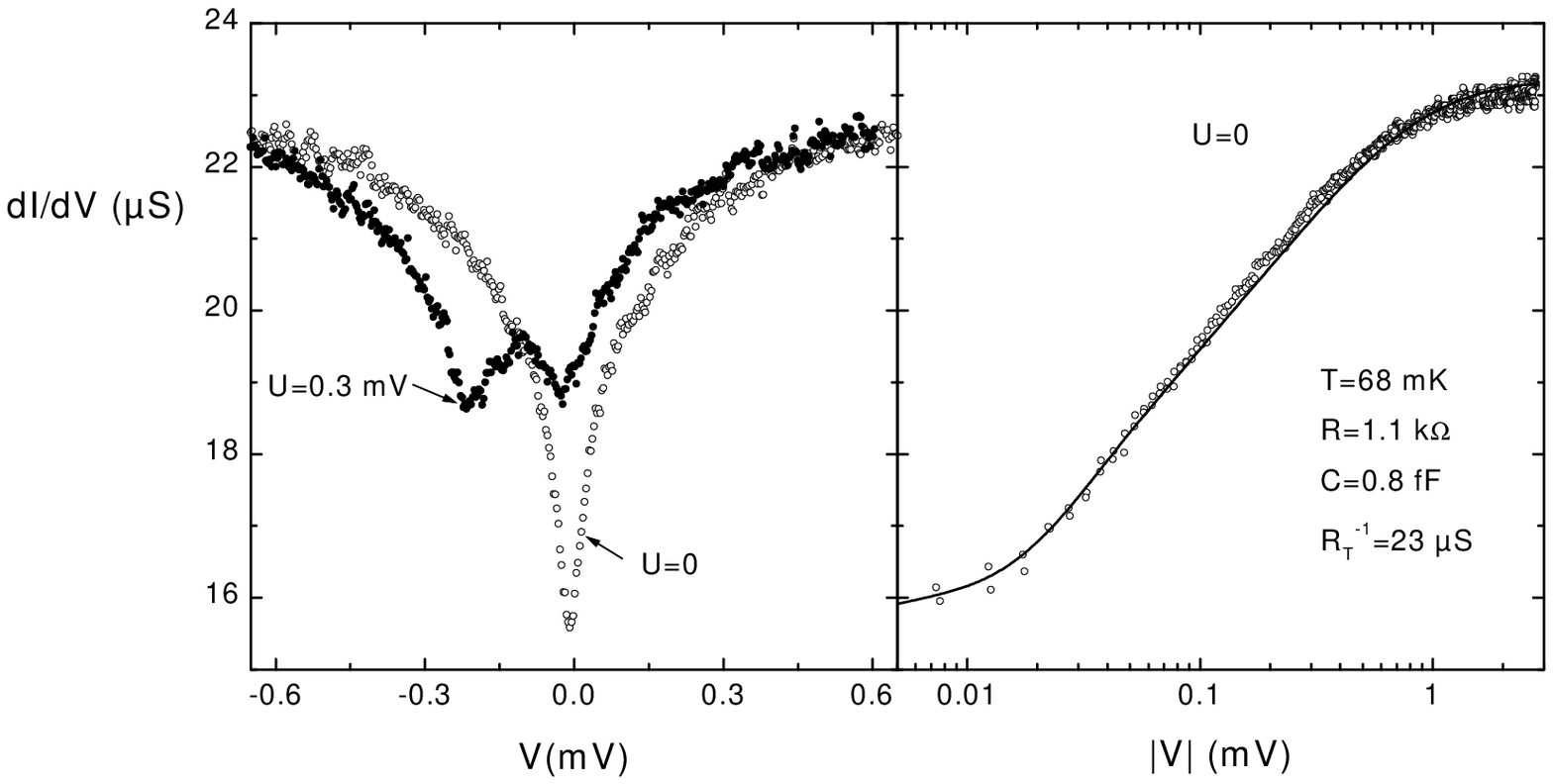}}
\caption{Left panel: Measured differential conductance at $B=0.8~\mathrm{T}$
for $U=0$ (open symbols) and $U=0.3~\mathrm{mV}$ (black symbols). Right panel: Data at $%
U=0$ reploted on a log-scale, together with the best fit with Eq. 1 (solid
line) obtained with $R_{\mathrm{T}}^{-1}=23~$\textrm{%
%TCIMACRO{\U{b5}}%
%BeginExpansion
$\mu$%
%EndExpansion
S}, 
$R=1.1~$\textrm{k}$\Omega $, $C=0.8~\mathrm{fF}$, and $T=68~$\textrm{mK.}}
\end{figure}

In a second step, the voltage $U$ is set to a finite value. We fit the
measured $\frac{dI}{dV}\left( V\right) $ with the calculated differential
conductance using Eq.~\ref{convolution1}. The function $P(\varepsilon )$
is calculated with the environmental
characteristics $(R, C, T)$ determined at $U=0$ and the fit parameters are the values of 
$f(E)$ at fourty regularly spaced energy values between $-eU-0.1~\mathrm{meV}
$ and $0.1~\mathrm{meV}$. Note that setting the temperature $T_\mathrm{ref}$ of the Fermi function $f_{\mathrm{ref}}(E)$ in the probe electrode also to $68~\mathrm{mK}$ causing numerical instabilities, the data shown here were obtained with $T_\mathrm{ref}=0$. 

\section{Experimental results}

\bigskip

\subsection{Distribution functions}

\bigskip

We show in Fig. 4 the distribution functions for two bias-voltages $U=0.1~%
\mathrm{mV}$ and $U=0.3~\mathrm{mV}$ and for magnetic fields varying from $%
0.16~\mathrm{T}$ to $2~\mathrm{T}$. Also shown with dotted lines are the 
zero-magnetic field curves, obtained with the probe in the superconducting state.
\begin{figure}[h]
\epsfysize=55mm \centerline{\epsfbox{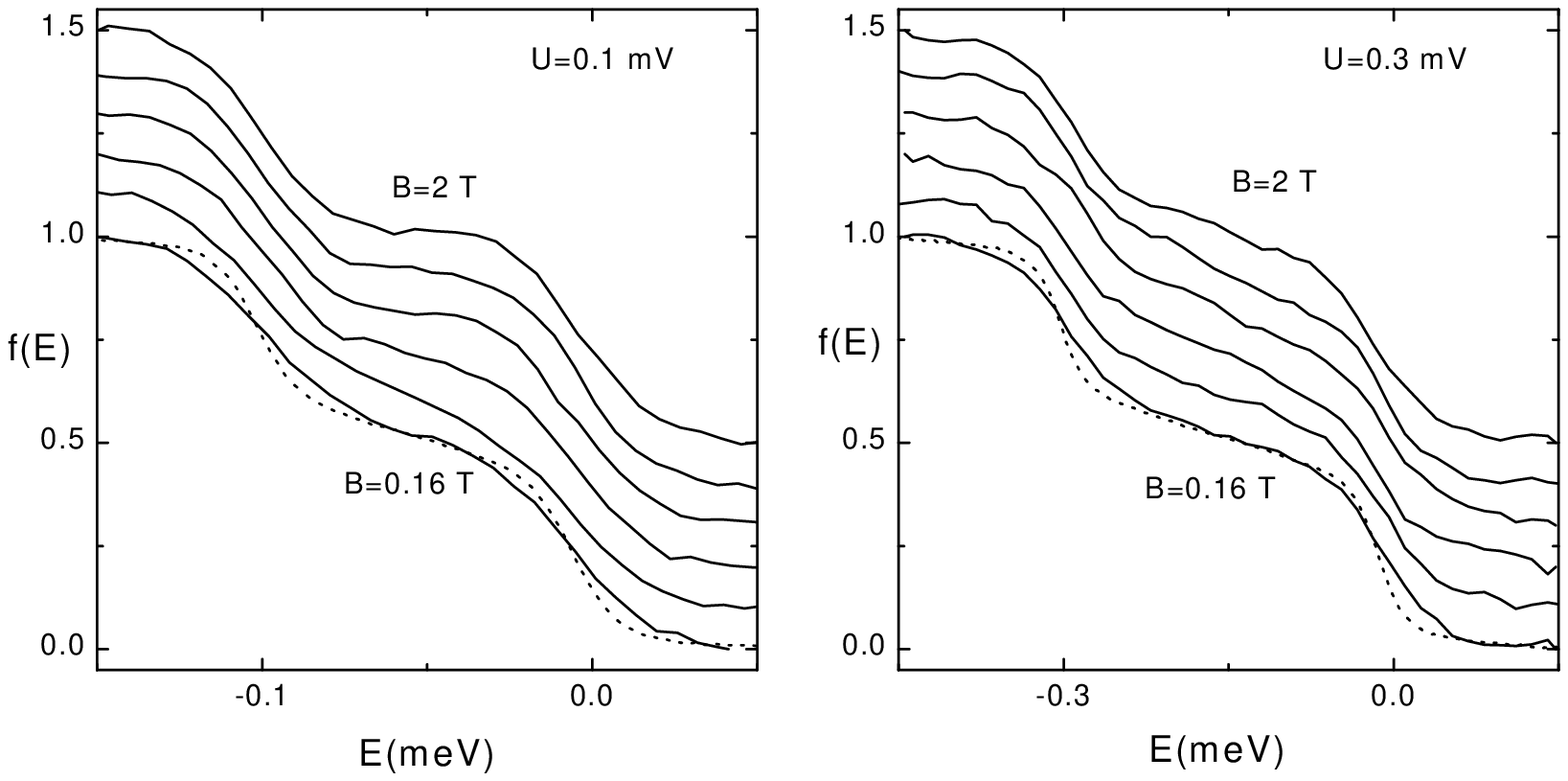}}
\caption{Measured distribution functions in the middle of the $5~$\textrm{%
%TCIMACRO{\U{b5}}%
%BeginExpansion
$\mu$%
%EndExpansion
m}-long copper wire. Solid line, from bottom to top: $B=0.16,\mathrm{~}%
0.4,~0.8,~1.2,~1.6,$ and$~2~\mathrm{T}$. Distribution functions are shifted
vertically by steps of $0.1$. Dotted line: Measured distribution function at 
$B=0$, obtained with the probe in the superconducting state.}
\end{figure}

In zero-magnetic field, we find that $f(E)$ only depends on $E/eU$. Such a
scaling property was found in all previous experiments on copper wires~\cite%
{relax,Fred}. In a magnetic field, scaling does not hold any more. Besides,
the evolution of $f(E)$ with the magnetic field is voltage-dependent: at $U=$
$0.1~\mathrm{mV}$, the plateau of the measured distribution functions near $%
f(E)=0.5$ first gets steeper with the magnetic field up to $0.4~\mathrm{T}$,
then flattens and is finally horizontal at $B=2~\mathrm{T}$; at $U=0.3~%
\mathrm{mV}$, it gets steeper as the magnetic field increases up to $B=1.6~%
\mathrm{T}$, then flattens again.

The excess rounding of the steps in a magnetic field compared to the zero-magnetic field data might be due to the choice $T_\mathrm{ref}=0$ in the deconvolution procedure, to an oversimplification of the environmental
impedance, to external noise, or, more fundamentally, to an increased energy exchange rate at low energies.

In the next section, we quantify the dependence of the interaction on $B$
and $U$.

\subsection{\protect\bigskip Energy exchange rates}

The distribution function can be calculated by solving the stationary
Boltzmann equation in the diffusive regime~\cite{Kozub,Nagaev}: 
\begin{equation}
\frac{1}{\tau _{D}}\frac{\partial ^{2}f\left( x,E\right) }{\partial x^{2}}+%
\mathcal{I}_{\mathrm{coll}}^{\mathrm{in}}\left( x,E,\left\{ f\right\}
\right) -\mathcal{I}_{\mathrm{coll}}^{\mathrm{out}}\left( x,E,\left\{
f\right\} \right) =0  \tag{2}  \label{Boltzmann}
\end{equation}%
where $\mathcal{I}_{\mathrm{coll}}^{\mathrm{in}}\left( x,E,\left\{ f\right\}
\right) $ and $\mathcal{I}_{\mathrm{coll}}^{\mathrm{out}}\left( x,E,\left\{
f\right\} \right) $ are the rates at which quasiparticles are scattered in
and out of a state at energy $E$ by inelastic processes. Assuming that the
dominant inelastic process is a two-quasiparticle interaction which is local
on the scale of variations of the distribution function, 
\begin{equation}
\mathcal{I}_{\mathrm{coll}}^{\mathrm{in}}\left( x,E,\left\{ f\right\}
\right) =\int \mathrm{d}\varepsilon \mathrm{d}E^{\prime }K\left( \varepsilon
\right) f_{E+\varepsilon }^{x}(1-f_{E}^{x})f_{E^{\prime
}}^{x}(1-f_{E^{^{\prime }}-\varepsilon }^{x})  \tag{3}  \label{Iout}
\end{equation}%
where the shorthand $f_{E}^{x}$ stands for $f\left( x,E\right) .$ The
out-collision term $\mathcal{I}_{\mathrm{coll}}^{\mathrm{out}}$ has a
similar form. The kernel function $K\left( \varepsilon \right) $ is
proportional to the averaged squared interaction between two quasiparticles
exchanging an energy $\varepsilon .$ We have neglected the possible
dependence of $K(\varepsilon )$ on the energies of the initial and final
states and on the position along the wire. The theory of Coulomb
interactions in disordered wires predicts $K(\varepsilon )\propto
\varepsilon ^{-3/2}$. However, in copper wires, at zero-magnetic field, the
scaling property implies, by a simple change of variables in Eq.{}~2, that $%
U^{2}K\left( \varepsilon \right) $ is a function of $\varepsilon /eU$ only~%
\cite{relax_1}.\ If futhermore $K\left( \varepsilon \right) $ does not
depend on $U,$ one obtains $K\left( \varepsilon \right) =\gamma /\varepsilon
^{2},$ with $\gamma $ a typical interaction rate, independent of $U$.

We have found that the curves $f(E)$ measured in a magnetic field could also
be reasonably fitted assuming $K\left( \varepsilon \right) =\gamma
(U,B)/\varepsilon ^{2}$ where $\gamma (U,B)$ is now a fit parameter for each
curve. It is only in zero-magnetic field that $\gamma (U,0)$ is independant
of $U$. We show in Fig. 5\ the evolution of this effective intensity $\gamma
(U,B)$ with the bias voltage and the magnetic field, in a grayscale plot. One observes that the typical interaction rate has a non-monotonous
behaviour with $B$. It increases roughly from $B=0$ to $B\approx \frac{eU}{%
4.3\mu _{B}}$ then decreases again. The zero-magnetic-field value $\gamma
(U,0)=0.52~\mathrm{ns}^{-1}$ is recovered near $B\approx \frac{eU}{2\mu _{B}}
$ (see straight line on Fig. 5).
\begin{figure}[h]
\epsfysize=57mm \centerline{\epsfbox{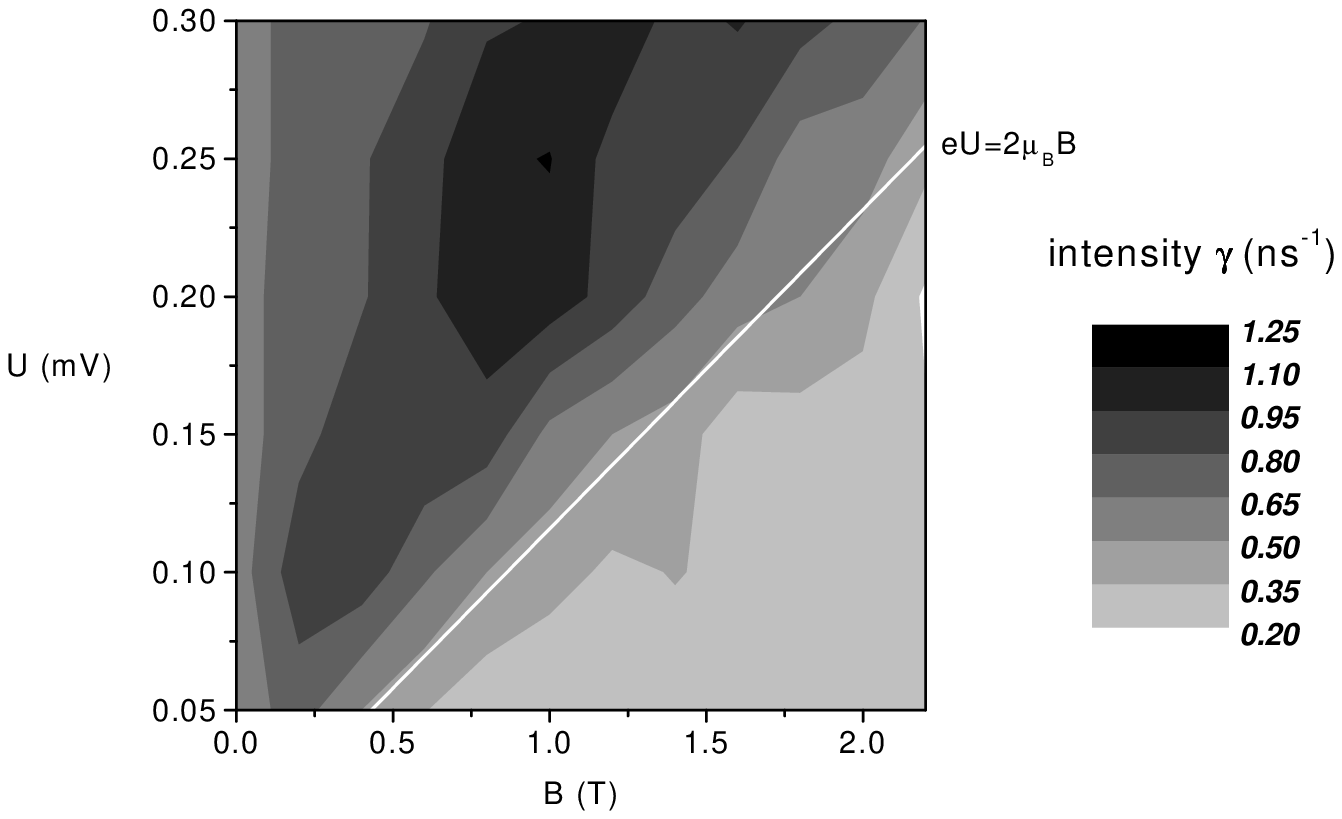}}
\caption{Evolution of the interaction intensity $\protect\gamma $ with the
magnetic field $B$ and the bias voltage $U$ obtained from a fit assuming $K(%
\protect\varepsilon )=\gamma /\varepsilon^{2}$. The
straight line is $eU=2%
%TCIMACRO{\U{b5}}%
%BeginExpansion
{\mu}%
%EndExpansion
_{B}B$.}
\end{figure}

\section{Magnetic-impurity-mediated energy exchange between electrons}

Kaminski and Glazman have proposed a second order process associated with
magnetic impurities in which energy is exchanged between quasiparticles~\cite%
{Glazman}. In the top of Fig.~6, we describe this second-order process
in the situation where, due to Zeeman effect, the spin-up and the spin-down
states of the magnetic impurities are split by the energy $g\mu _{B}B$.\ For
simplicity, the impurities are assumed to be spin $1/2$; $g$ is the
gyromagnetic factor of the impurities. In Fig. 6, the initial polarisation
of the impurity is assumed to correspond to its lowest-energy state. The
energy of the intermediate state is then $-\varepsilon +g%
%TCIMACRO{\U{b5}}%
%BeginExpansion
{\mu}%
%EndExpansion
_{B}B$, so that the rate of this process of second order in perturbation is
proportional to $(-\varepsilon +g%
%TCIMACRO{\U{b5}}%
%BeginExpansion
{\mu}%
%EndExpansion
_{B}B)^{-2}$. At $B=0,$ one obtains $K(%
\protect\varepsilon )\propto \varepsilon^{-2}$.

On the other hand, in a magnetic field, inelastic scattering also occurs in
the first order process: the energy $\pm g\mu _{B}B$ is exchanged directly
between an electron and a magnetic impurity (see Fig. 6, bottom). The
magnetic impurities then behaves as two-level systems, the occupation of the
two states being determined self-consistently, at each position along ther
wire, by the coupling to the quasiparticles \cite{relaxAg}. This process
could explain why the interaction rate $\gamma $ first increases with 
magnetic field.
\begin{figure}[h]
\epsfysize=82mm \centerline{\epsfbox{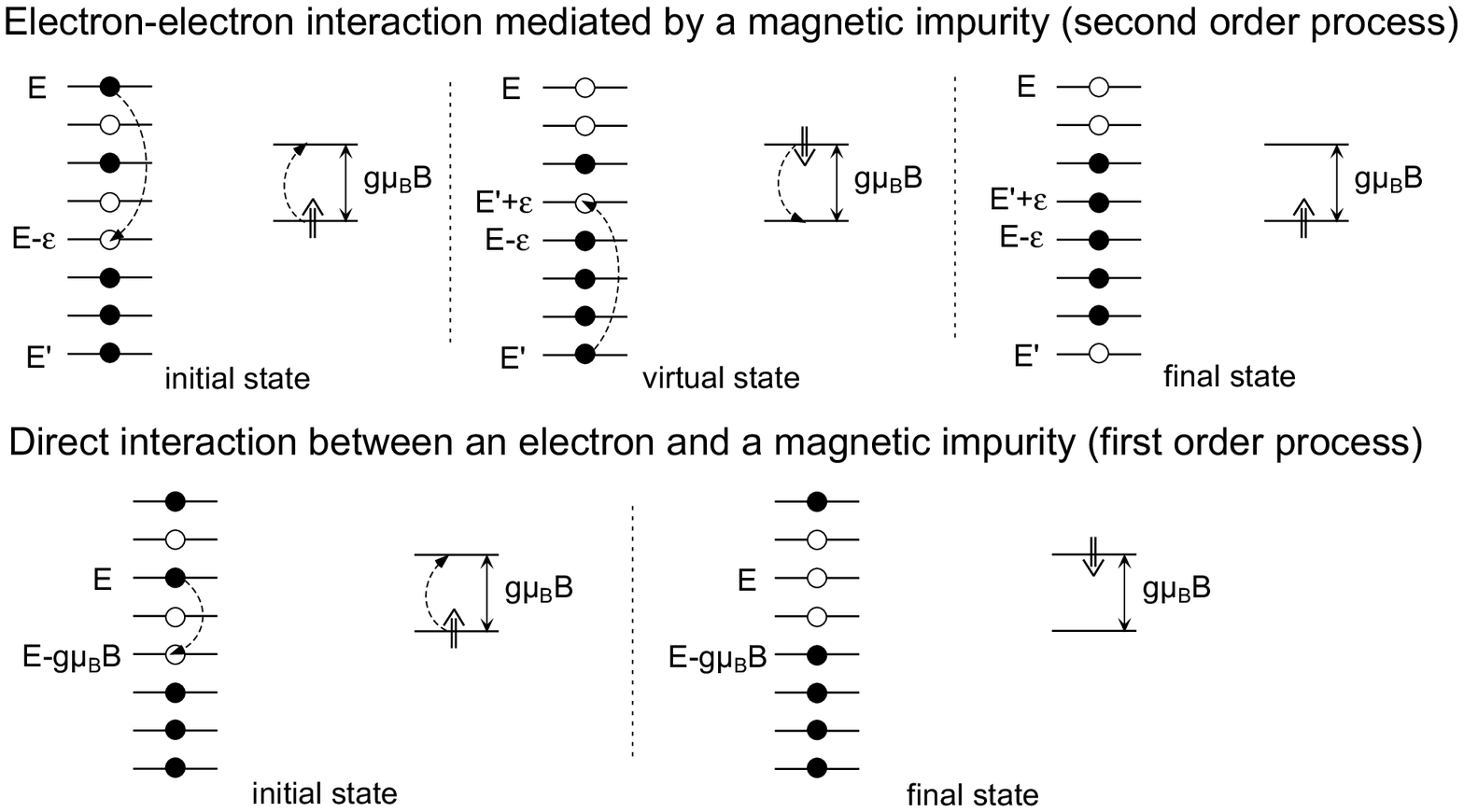}}
\caption{Description of the processes of redistribution of energy between
quasiparticles mediated by magnetic impurity. In each panel, the left ladder
represents the energy spectrum of the quasiparticles and the two states on
the right side represent the energy levels of the spin-up and spin-down
state of a magnetic impurity. Top: The second order process implies two
electrons and the spin of a magnetic impurity. Bottom: The first order
process directly exchanges $g%
%TCIMACRO{\U{b5}}%
%BeginExpansion
{\mu}%
%EndExpansion
_{B}B$ between an electron and the spin of a magnetic impurity.}
\end{figure}

If the splitting $g\mu _{B}B$ becomes larger than $eU$, starting from an
occupied electron-state at energy $E$, the state of energy $E-g\mu _{B}B$ is
occupied. Then, due to Pauli principle, the rate for first order processes 
vanishes, and the magnetic
impurities remain in their ground state. 
The rate of the second-order processes also vanishes at large magnetic fields 
because of the $g\mu _{B}B$ term in its denominator and of the Pauli constraint 
$\varepsilon \lesssim eU$. This explains the
decrease of the interaction rate at $B\approx \frac{eU}{g\mu _{B}}$.

A quantitative description of these processes, taking into account the
renormalisation of the coupling between quasiparticles and magnetic
impurities due to Kondo effect, is under progress~\cite{Georg2}. 

\newpage
In conclusion, we have found that the anomalous energy exchange rate between
quasiparticles in copper wire strongly depends on magnetic field. The
magnetic field and bias-voltage dependence of the intensity of the
interaction are in qualitative agreement with a magnetic-impurity-mediated
interaction. The nature of the magnetic impurities is not known, but
measurements of the phase-coherence time give evidence that the copper oxyde
might play a role~\cite{Haesendonck}.

\section*{Acknowledgments}

We are grateful to G. G\"{o}ppert, H. Grabert, T. Giamarchi and A. Georges
for useful discussions and comments, P. Joyez for permanent assistance, 
and P.F. Orfila for technical support. This work was partially supported by the Bureau National de M\'etrologie.


\section*{References}

\begin{thebibliography}{99}
\bibitem{Echternach} S.\ Wind, M.J.\ Rooks, V.\ Chandrasekhar, and D.E.\
Prober, Phys. Rev.\ Lett. \textbf{57}, 633 (1986); P.M. Echternach, M.E.
Gershenson, H.M. Bozler, A.L. Bogdanov and B. Nilsson, \textit{Phys. Rev. }%
\textbf{B} \textbf{48}, 11516 (1993); P. Mohanty, E.M.Q. Jariwala and R.A.
Webb, \textit{Phys. Rev. Lett.} \textbf{78}, 3366 (1997).

\bibitem{WLGoteborg} A.B.\ Gougam, F.\ Pierre, H.\ Pothier, D.\ Esteve, and
N.O.\ Birge, \emph{J. Low Temp. Phys.} \textbf{118}, 447 (2000).

\bibitem{AA} For a review, see B.L.\ Altshuler and A.G.\ Aronov, in \emph{%
Electron-Electron Interactions in Disordered Systems,} Ed.\ A.L.\ Efros and
M.\ Pollak (Elsevier Science Publishers B.V., 1985).

\bibitem{relaxAg} F.\ Pierre, H. Pothier, D. Esteve, and M. H. Devoret, 
\emph{J. Low Temp. Phys.} \textbf{118}, 437 (2000)

\bibitem{relax} H. Pothier, S. Gu\'{e}ron, Norman O. Birge, D. Esteve, and
M. H. Devoret, \emph{Phys. Rev. Lett.} \textbf{79}, 3490 (1997).

\bibitem{Fred} F. Pierre, thesis, Universit\'{e} Paris 6, Paris, 2000.

\bibitem{Pesc} F.\ Pierre, H.\ Pothier, D.\ Esteve, M.\ H.\ Devoret, A. B.\
Gougam, and Norman O.\ Birge, Proceedings of the NATO Advanced Research
Workshop on \emph{Size Dependent Magnetic Scattering}, Ed. V. Chandrasekhar
and C. Van Haesendonck (2001). (cond-mat/0012038)

\bibitem{Glazman} A. Kaminski and L. I. Glazman, \emph{Phys. Rev. Lett.} 
\textbf{86}, 2400 (2001).

\bibitem{georg} G. G\"{o}ppert and H. Grabert, \emph{Phys. Rev. }\textbf{B} 
\textbf{64}, 033301 (2001).

\bibitem{Kroha} J.\ Kroha and A.\ Zawadowski, cond-mat/0104151.

\bibitem{Steinbach} A. H.\ Steinbach, J. M.\ Martinis, and M. H.\ Devoret, 
\emph{Phys. Rev. Lett.} \textbf{76}, 3806 (1996).

\bibitem{Kozub} V. I.\ Kozub and A.\ M.\ Rudin, \emph{Phys. Rev. }\textbf{B} 
\textbf{52}, 7853 (1995).

\bibitem{Devoret} For a review, see G.-L.\ Ingold and Yu.\ Nazarov, in \emph{%
Single Charge Tunneling,} Ed. H. Grabert and M. H.\ Devoret (Plenum
Press, New York, 1992).

\bibitem{filtres} D.\ Vion, P.F.\ Orfila, P.\ Joyez, D.\ Esteve, and M. H.\
Devoret, \emph{J.\ Appl. Phys.} \textbf{77}, 2519 (1995).

\bibitem{Nagaev} K.E.\ Nagaev, \emph{Phys. Lett.\ A} \textbf{169}, 103
(1992); \emph{Phys. Rev. }\textbf{B} \textbf{52}, 4740 (1995).

\bibitem{relax_1} H. Pothier, S. Gu\'{e}ron, Norman O. Birge, D. Esteve, and
M. H. Devoret, \emph{Z.\ Phys. }\textbf{B} \textbf{104}, 178 (1997).

\bibitem{Georg2} G. G\"{o}ppert, private communication.

\bibitem{Haesendonck} J.\ Vranken, C.\ Van Haesendonck, and Y.\
Bruynseraede, \emph{Phys. Rev. }\textbf{B} \textbf{37}, 8502 (1988).
\end{thebibliography}
\end{document}